\documentclass[superscriptaddress,secnumarabic,twocolumn,
 amssymb,amsmath,nobibnotes,aps,prd,showkeys,showpacs,nofootinbib]{revtex4}
\usepackage{graphicx,epsfig,amssymb,amsmath,amstext}
\usepackage{bm}
\usepackage{booktabs}
\usepackage{color}
 
\let\baraccent=\= 
\renewcommand{\=}[1]{\stackrel{#1}{=}} 

\begin{document}
\title{Time parameterizations and spin supplementary conditions of the 
Mathisson-Papapetrou-Dixon equations}

\author{Georgios Lukes-Gerakopoulos}
\email{gglukes@gmail.com}
\affiliation{Astronomical Institute of the Academy of Sciences of the Czech Republic,
Bo\v{c}n\'{i} II 1401/1a, CZ-141 31 Prague, Czech Republic}

\begin{abstract}
 The implications of two different time constraints on the
 Mathisson-Papapetrou-Dixon (MPD) equations are discussed under three spin
 supplementary conditions (SSC). For this reason the MPD equations are revisited
 without specifying the affine parameter and several relations are reintroduced
 in their general form. The latter allows to investigate the consequences of
 combining the Mathisson-Pirani (MP) SSC, the Tulczyjew-Dixon (TD) SSC and the
 Ohashi-Kyrian-Semer\'{a}k (OKS) SSC with two affine parameter types:
 the proper time on one hand and the parameterizations introduced in
 [{\it Gen. Rel. Grav.} {\bf 8}, 197 (1977)] on the other. 
 For the MP SSC and the TD SSC it is shown that quantities that are constant of
 motion for the one affine parameter are not for the other, while for the OKS
 SSC it is shown that the two affine parameters are the same.
 To clarify the relation between the two affine parameters in the case of the
 TD SSC the MPD equations are evolved and discussed. 
\end{abstract}
%
%

\maketitle
\tableofcontents{}

\section{Introduction}\label{sec:Intro} 

 The motion of a small mass body whose effect on the spacetime background is
 negligible had been first studied in terms of the multipole moments of the body
 by Mathisson \cite{Mathisson37} and Papapetrou \cite{Papapetrou51}. A covariant 
 formalism was achieved by Dixon in \cite{Dixon64}, who also reformulated the
 respective equations of motion. These equations of motion are known now as
 Mathisson-Papapetrou-Dixon (MPD) equations.
 
 In the case of a solely gravitational interaction within the pole-dipole
 approximation the MPD equations read
 \begin{eqnarray}
  \dot{p}^\mu=-\frac{1}{2}~{R^\mu}_{\nu\kappa\lambda} v^\nu S^{\kappa\lambda}\quad,
  \label{eq:MPmomenta} \\
  \nonumber \\
  \dot{S}^{\mu\nu}=p^\mu~v^\nu-v^\mu~p^\nu\equiv 2 p^{[\mu}v^{\nu]} \quad, \label{eq:MPspin}
 \end{eqnarray}
 where $p^\mu$ is the four-momentum, $v^\mu=d x^\mu/d \chi$ is the tangent
 vector and $S^{\mu\nu}$ is the spin tensor of the body. Moreover, 
 ${R^\mu}_{\nu\kappa\lambda}$ is the Riemann tensor and the dot denotes a
 covariant differentiation along the worldline $x^\mu(\chi)$, where $\chi$ is an
 evolution parameter along the worldline not necessarily the proper time $\tau$.
 Thus, it is not assumed that the tangent vector is the four-velocity, and the
 contraction 
 \begin{align}\label{eq:4VelCon}
 v^\mu v_\mu \equiv -v^2
 \end{align}
 does not represent necessarily the four-velocity preservation $v^2=1$. 

 The notion of mass can be defined either with respect to the momentum $p^\nu$,
 i.e.
 \begin{align} \label{eq:MomMass}
  m^2 \equiv-p^\nu~p_\nu   \quad,
 \end{align}
  or with respect to the tangent vector $v^\nu$, i.e.
  \begin{align} \label{eq:VelMass}
  m_v\equiv-v^\nu~p_\nu \quad.
 \end{align}
 
 \paragraph*{Units and notation:}
  The units  employed in this work are geometric $(G=c=1)$, and the signature of
 the metric $g_{\mu\nu}$ is (-,+,+,+). Greek letters denote the indices
 corresponding to spacetime (running from 0 to 3). The Riemann tensor
 is defined as
${{R^\alpha}_{\beta\gamma\delta}=
 \Gamma^\alpha_{\gamma \lambda} \Gamma^\lambda_{\delta \beta}
 - \partial_\delta \Gamma^\alpha_{\gamma\beta}
 - \Gamma^\alpha_{\delta\lambda} \Gamma^{\lambda}_{\gamma\beta}
 + \partial_\gamma \Gamma^{\alpha}_{\delta \beta}}$,
where $\Gamma$ are the Christoffel symbols. The Levi-Civita tensor is
$\epsilon_{\kappa\lambda\mu\nu}=\sqrt{-g} \tilde{\epsilon}_{\kappa\lambda\mu\nu} $
with the Levi-Civita symbol defined as $\tilde{\epsilon}_{0123}=1$.

 \section{Some useful relations} \label{sec:UsRe}
 
 By keeping in mind that $v^2$ is not necessarily constant, some useful 
 consequences of MPD equations are presented below. These consequences coincide
 with expressions presented in \cite{Semerak99} when $v^2=1$. 
 
 Contracting Eq.~\eqref{eq:MPmomenta} with $v_\mu$
 gives
 \begin{align} \label{eq:UC1}
  \dot{p}^\mu v_\mu=0 \quad.
 \end{align}
 Contraction of Eq.~\eqref{eq:MPspin} with $v_\mu$ gives
  \begin{align} \label{eq:UC2}
   p^\mu=\frac{1}{v^2}(m_v~v^\mu-\dot{S}^{\mu\nu} v_\nu) \quad,
  \end{align}
 while contracting with $\dot{v}_\mu$ and using relation~\eqref{eq:UC2} gives
 \begin{align} \label{eq:UC3}
  \dot{v}_\mu \dot{S}^{\mu\nu} =\frac{\dot{v}_\mu}{v^2} (v^\mu \dot{S}^{\nu\rho}-\dot{S}^{\mu\rho} v^\nu) v_\rho \quad.
 \end{align}
 Contracting Eq.~\eqref{eq:MPspin} with $p_\mu$ gives
  \begin{align} \label{eq:UC4}
   p^\mu=\frac{1}{m_v}(m^2~v^\mu-\dot{S}^{\mu\nu} p_\nu) \quad,
  \end{align}
 while contracting with $\dot{p}_\nu$ and using relation~\eqref{eq:UC4} gives 
  \begin{align} \label{eq:UC5}
    \dot{p}_\nu \dot{S}^{\mu\nu} =\frac{1}{m_v} \dot{p}_\nu \dot{S}^{\nu\rho} p_\rho v^\mu \quad.
  \end{align}
 Contracting Eq.~\eqref{eq:UC4} with $v^\mu$ leads to
  \begin{align} \label{eq:UC6}  
    m_v^2- m^2 v^2=v_\mu \dot{S}^{\mu\nu} p_\nu \quad,
  \end{align} 
  which combined with Eq.~\eqref{eq:UC2} gives
  \begin{align} \label{eq:UC7}
     m^2 v^4-m_v^2 v^2=v^\kappa \dot{S}_{\kappa\mu} \dot{S}^{\mu\nu} v_\nu \quad.
  \end{align}

  Furthermore, one finds that the evolution equation of the mass $m$ is
  \begin{align} \label{eq:MomMassEvol}
   \dot{m}=-\frac{1}{m}\dot{p}_\mu p^\mu=\frac{1}{m~m_v}\dot{p}_\mu \dot{S}^{\mu\nu} p_\nu \quad,
   \end{align}
  for which result Eqs.~\eqref{eq:UC4},~\eqref{eq:UC5} are used, while the
  evolution equation of the mass $m_v$ is 
  \begin{align} \label{eq:VelMassEvol}
   \dot{m}_v=-\frac{1}{v^2}(v_\mu \dot{S}^{\mu\nu}+m_v v^\nu)\dot{v}_\nu \quad,
  \end{align}
  for which result Eq.~\eqref{eq:UC4} is used, and Eq.~\eqref{eq:UC1} is taken
  into account. 
  
  The square of the spin's measure is 
  \begin{align} \label{eq:SpinMeas}
    S^2=\frac{1}{2}~S_{\mu\nu}~S^{\mu\nu} \quad,
  \end{align}  
  and its evolution equation reads
 \begin{align} \label{eq:SpinMeasEvol}
   \dot{S}^2 = 2 p_\mu S^{\mu\nu} v_\nu \quad,
 \end{align} 
 in which calculation Eq~\eqref{eq:MPspin} is used.
   
 \section{Choosing a worldline}

  The MPD equation system, consisted of \eqref{eq:MPmomenta},~\eqref{eq:MPspin}
  and $d x^\mu/d \chi=v^\mu$,  is under-defined. Namely, there are only 14
  independent equations of motion for the 18 variables
  $\{x^\mu,v^\mu,p^\mu,S^{\mu\nu} \}$\footnote{Because the spin tensor $S^{\mu\nu}$
  is antisymmetric, it contributes only 6 independent equations and variables.}.
  To define a worldline we have to supplement the system with 4 additional
  constraints.
  
  One of these constraints comes from choosing the evolution parameter $\chi$.
  A common choice for the evolution parameter is to identify $\chi$ with the
  proper time $\tau$, see, e.g., \cite{Dixon64,Semerak99}. Then,
  $v^2=1$ and the tangent vector $v^\mu$ identifies with the four-velocity. 
  Another interesting choice was introduced in \cite{Ehlers77}, according to
  which $\chi$ scales in such way that 
  \begin{align} \label{eq:Ehlers}
   v_\mu u^\mu =-1 \quad,
  \end{align}
  where $u^\mu=p^\mu/m$. An apparent consequence of this choice is that
  $m_v=m$. This affine parameter is denoted as $\sigma$.
  
  After having chosen the evolution parameter, the remaining necessary constraints
  are devoted to choose the center of the mass of the system. The center of
  mass is called often centroid. By choosing the centroid and the evolution 
  parameter one defines the evolution along the worldline that the body
  described by the MP equations follows. In particular, the centroid is fixed by
  choosing an observer through a time-like vector $V^\mu$ for which
  $V_\mu S^{\mu\nu}=0$. This constraint is known as spin supplementary condition
  (SSC). In the bibliography there are five established choices of SSC:
  \begin{enumerate}
   \item the Mathisson-Pirani (MP) condition $V^\mu=v^\mu$ \cite{Mathisson37,Pirani56}.
   \item the Tulczyjew-Dixon (TD) condition  $V^\mu=u^\mu$ \cite{Tulczyjew59,Dixon1970I}.
   \item the Corinaldesi-Papapetrou condition  $V^\mu = v_\textrm{lab}$ \cite{Corinaldesi51},
   where $v_\textrm{lab}$ is a a congruence of ``laboratory'' observers.
   \item the Newton-Wigner condition $V^\mu \propto v_\textrm{lab}+u^\mu$ \cite{NewtonWigner49}.
   \item the Ohashi-Kyrian-Semer\'{a}k (OKS) condition  \cite{Ohashi03,Kyrian07}),
   for which the $V^\mu$ is chosen in such way that $p^\mu\parallel v^\mu$.
  \end{enumerate}

   \section{Discussing the constraints}
   
   In this section are investigated the consequences of combining the time
   constraints and particularly the condition~\eqref{eq:Ehlers} with the MP SSC,
   the TD SSC and the OKS SSC. For these three SSCs Eq.~\eqref{eq:SpinMeasEvol}
   shows that the measure of the spin is conserved independently from the time
   constraint choice.
   
   \subsection{The Mathisson-Pirani SSC} \label{sec:MPSSC}   
   
   Contracting the covariant derivative of MP SSC with $\dot{v}_\nu$ results
   in $v_\mu \dot{S}^{\mu\nu} \dot{v}_\nu=0$, taking this into account
   Eq.~\eqref{eq:VelMassEvol} for the MP SSC gives
   \begin{align} \label{eq:MPERcon}
   \frac{\dot{m}_v}{m_v}=\frac{ \dot{v^2}}{2 v^2} \Rightarrow \frac{m_v^2}{v^2}=const. \quad.
   \end{align}
   For the condition $v^2=1$, Eq.~\eqref{eq:MPERcon} gives that $m_v$
   is a constant of motion. For the condition~\eqref{eq:Ehlers}, 
   Eq.~\eqref{eq:MPERcon} gives that 
   \begin{align}
    \frac{m^2}{v^2}=const.\quad,
   \end{align}
   since $m=m_v$.
   
   \subsection{The Tulczyjew-Dixon SSC} \label{sec:TDSSC}
   
   For TD SSC Eq.~\eqref{eq:MomMass} shows that the mass $m$ is a constant of
   motion. Since for the condition~\eqref{eq:Ehlers} $m=m_v$, then $m_v$ is
   constant as well. Thus, Eq.~\eqref{eq:VelMassEvol} gives 
   \begin{align}\label{eq:VelMassCon}
    (v_\mu \dot{S}^{\mu\nu}+m_v v^\nu)\dot{v}_\nu=0 \quad.
   \end{align}
   
   According to Eq.~\eqref{eq:VelMassCon}, if $v^\nu \dot{v}_\nu=0$, then it holds
   that $v_\mu \dot{S}^{\mu\nu} \dot{v}_\nu=0$ as well. The former implies
   that $v^2$ is a constant, while the latter implies that MP SSC holds along
   with the assumed TD SSC. The last implication is proven as follows:
   contracting Eq.~\eqref{eq:UC3} with $v_\nu$ gives $\dot{S}^{\mu\rho} v_\rho=0$,
   because $v_\mu \dot{S}^{\mu\nu} \dot{v}_\nu=0$ and it is reasonable to assume 
   that $v^2=0$ is not the case. Since $v^2$ and $m_v$ are constants
   and it has been shown that $\dot{S}^{\mu\rho} v_\rho=0$, Eq.~\eqref{eq:UC2}
   results in $p^\mu||v^\mu$. If $p^\mu||v^\mu$, then Eq.~\eqref{eq:UC6} gives
   $v^2=1$. Therefore, when $v^\nu \dot{v}_\nu=0$ the affine parameter
   defined by the condition~\eqref{eq:Ehlers} is the proper time, i.e.
   $\sigma=\tau$.
   
   The cases of TD SSC for which $p^\mu||v^\mu$ holds are very special cases.
   Thus, it is reasonable to assume that for the condition~\eqref{eq:Ehlers}   
   in general $v^\nu \dot{v}_\nu \neq 0$ is true. Under this assumption,
   Eq.~\eqref{eq:VelMassCon} gives
   \begin{align}
    m_v= -\frac{v_\mu \dot{S}^{\mu\nu} \dot{v}_\nu}{v^\nu \dot{v}_\nu}=const. \quad.
   \end{align}
   In this case Eq.~\eqref{eq:UC6} implies that the variation of $v^2$ during
   the evolution is reflected on the $v_\mu \dot{S}^{\mu\nu} p_\nu$ evolution.
   Actually, if one uses the $v^2=1$ condition instead of the
   condition~\eqref{eq:Ehlers}, then the variation of $m_v^2$ during
   the evolution is reflected on the $v_\mu \dot{S}^{\mu\nu} p_\nu$ evolution.
   
   Another interesting relation comes from eq.~\eqref{eq:UC6}, when one uses
   the covariant derivative of the TD SSC and then applies eq.~\eqref{eq:MPmomenta},
   then we get
  \begin{align} \label{eq:UC6TD}  
    v^2=\frac{1}{m^2}(m_v^2-\frac{1}{2} v_\sigma S^{\sigma\mu} R_{\mu\nu\rho\alpha} v^\nu S^{\rho\alpha}) \quad.
  \end{align}

   \subsection{The Ohashi-Kyrian-Semer\'{a}k  SSC} \label{sec:OKSSSC}
   
   Since for OKS SSC by definition $p^\mu||u^\mu$, then it holds that
   $v_\mu \dot{S}^{\mu\nu} p_\nu=0$. Combing the latter with the fact that
   $m=m_v$ for the condition~\eqref{eq:Ehlers},  Eq.~\eqref{eq:UC6} gives that
   $v^2=1$. This means that OKS SSC is satisfying both time constraints
   simultaneously; or in other words the affine parameter $\sigma$ is identical
   with the proper time $\tau$ for OKS SSC. Note that the latter holds when
   $(p^\mu||u^\mu)$ independently of the implemented SSC.

  \section{Numeric comparison for Tulczyjew-Dixon SSC}
  
   This section examines the evolution of the MPD equations under the two time 
   choices $\tau$ and $\sigma$. In particular, we are going to examine numerically
   the MPD under the TD SSC, since for OKS SSC the two evolution parameter
   choices are equivalent and for MP SSC the helical motion introduces an
   unnecessary complication.
   
   \subsection{Preliminary considerations}\label{sec:Prel}
   
   To do a numerical comparison, the first issue is the initial condition setup,
   i.e. the initial position, momentum and spin tensor have to be properly chosen.
   Since we have the same SSC (TD SSC), we have two observers with initially
   the same position $x^\mu$. The definitions of the momentum and the spin
   tensor depend only on the position $x^\mu$ along the worldline
   \cite{Dixon1970I}, hence the initial conditions $x^\mu$, $p_\mu$ and $S^{\mu\nu}$
   for two observers are the same. 
  
   However, the two observers are equipped with clocks that do not tick the same,
   i.e. they follow different affine parameters. If the  momentum and the spin
   tensor are not affected by the different choices of the affine parameter,
   then a time reparametrization of the MPD
   equations~\eqref{eq:MPmomenta}-\eqref{eq:MPspin} just means that the 
   MPD equations will reproduce the same worldline under different affine
   parameter $\chi$. The validity of the last statement is what is in this
   section checked.

   To evolve the MPD with TD SSC, one needs the relation
   \begin{align} \label{eq:v_p_TUL}
    v^\mu = \frac{m_v}{m^2} \left(
          p^\mu + 
          \frac{ 2 \; S^{\mu\nu} R_{\nu\rho\kappa\lambda} p^\rho S^{\kappa\lambda}}
          {4~m^2 + R_{\alpha\beta\gamma\delta} S^{\alpha\beta} S^{\gamma\delta} }
          \right)  \quad ,
   \end{align}
   which gives $v^\mu$ as function of $x^\mu$, $p_\mu$ and $S^{\mu\nu}$. An
   interesting fact about relation~\eqref{eq:v_p_TUL} is that its derivation does
   not depend on the time constraint (see, e.g., \cite{Ehlers77,Semerak99} for 
   the derivation). A related fact is that the relation~\eqref{eq:v_p_TUL}
   is invariant under affine parameter changes, since the scalar $m_v$ contains 
   the tangent vector $v^\mu$ (definition~\eqref{eq:VelMass}).
   
   The background, on which the MPD are to be evolved, is the Kerr spacetime.
   The metric tensor of Kerr in Boyer-Lindquist (BL) coordinates
   $\{t,r,\theta,\phi\}$ reads
   \begin{align}\label{eq:KerrMetric}
    g_{tt} &=-1+\frac{2 M r}{\Sigma} \quad , \quad g_{t\phi} = -\frac{2 a M r \sin^2{\theta}}{\Sigma}\quad, \nonumber\\
    g_{\phi\phi} &= \frac{\Lambda \sin^2{\theta}}{\Sigma} \quad , \quad g_{rr} = \frac{\Sigma}{\Delta}\quad,
    \quad g_{\theta\theta} = \Sigma \quad , 
   \end{align} 
   where
   \begin{align} \label{eq:Kerrfunc} 
    \Sigma &= r^2+ a^2 \cos^2{\theta}\quad, \quad  \Delta = \varpi^2-2 M r\quad,\nonumber \\ 
    \varpi^2 &= r^2+a^2\quad, \quad  \Lambda = \varpi^4-a^2\Delta \sin^2\theta\quad. 
   \end{align}
   and $M$ defines the mass and $a$ the Kerr parameter. The motion of a small
   spinning body in the stationary and axisymmetric Kerr spacetime preserves,
   respectively, the energy
   \begin{align}
    E &= -p_t+\frac12g_{t\mu,\nu}S^{\mu\nu}\quad,\label{eq:Energy}
   \end{align}
   and the component of the total angular momentum along the symmetry axis $z$ 
   \begin{align}
    J_z &= p_\phi-\frac12g_{\phi\mu,\nu}S^{\mu\nu}\quad.\label{eq:TotAnglularZ}
   \end{align}

   The MPD equations are valid when the size of the spinning body is much smaller
   than the curvature, i.e. when
   $$\lambda=\frac{|R_{\mu\nu\kappa\lambda}|}{\rho^{2}}\ll 1 \quad,$$
   where $ |R_{\mu\nu\kappa\lambda}|$ is the magnitude of the Riemann tensor and
   $\rho$ is the radius of the spinning body. If the radius $\rho$ is
   approximated by the M\o{}ller radius \cite{Moller49}, then $\rho=S/m$.
   Thus, since $ |R_{\mu\nu\kappa\lambda}|\sim M/r^3$, we get
   \begin{align}\label{eq:MPD_valid}
    \lambda \sim\left(\frac{S}{m~M}\right)^2\left(\frac{M}{r}\right)^3\quad.
   \end{align}  
   For the computations of the MPD equations, the dimensionless counterparts of
   the involved quantities are employed. For example, in theirs dimensionless
   forms the spin of the small body reads $S/(m M)$, the BL radius reads $r/M$
   and for the momentum holds that $p^\mu=u^\mu$. Numerically the values of
   dimensionless quantities are equal to the dimensionful by setting $M=m=1$.
   From this point on in the article there is no distinction between the  
   dimensionful and the dimensionless quantities. Since our discussion is
   theoretical, the spin of the small body does not need to be very
   small\footnote{See, e.g., Ref.~\cite{Hartl03} for thorough discussion the 
   astrophysically relevant spin values} as long as $\lambda \ll 1$. For a 
   radius $r \sim 10$ and $\lambda \sim 10^{-3}$, 
   approximation~\eqref{eq:MPD_valid} gives that $S\sim 1$. It is advantageous
   to use large spin values, because the larger the value is, the greater
   might be the difference between $v^\mu$ and $u^\mu$, and consequently the
   divergence between the two time constraints. 
   
   Following the initial condition setup presented in \cite{Hartl03}, instead of
   the spin tensor $S^{\mu\nu}$ the spin four-vector 
   \begin{align} \label{eq:SpinVect}
     S_\mu \equiv -\frac{1}{2} \epsilon_{\mu\nu\rho\sigma}
          \, u^\nu \, S^{\rho\sigma} 
   \end{align}
   is utilized. According to Ref.~\cite{Hartl03} setup, one can set $t=\phi=0$
   and provide the initial values for $r,~\theta,~u^r,~S^r,~S^\theta$, while the
   rest of the initial conditions $u^t$, $u^\theta$, $u^\phi$, $S^t$, and $S^\phi$,
   are fixed by $m$ (Eq.~\eqref{eq:MomMass}), $S$ (Eq.~\eqref{eq:SpinMeas}),
   $E$ (Eq.~\eqref{eq:Energy}), $J_z$ (Eq.~\eqref{eq:TotAnglularZ}) and the
   constraint $u_\mu S^\mu=0$. The latter constraint is obtained from contracting
   Eq.~\eqref{eq:SpinVect} with $u^\mu$, while in
   Eqs.~\eqref{eq:SpinMeas},~\eqref{eq:Energy},~\eqref{eq:TotAnglularZ}
   the inverse relation of Eq.~\eqref{eq:SpinVect} 
   \begin{equation}\label{eq:T4VSin}
    S^{\rho\sigma}=-\eta^{\rho\sigma\gamma\delta} S_{\gamma} u_\delta 
   \end{equation}
   is employed.
   
   \subsection{Numerical results}\label{sec:NumRes}
   
  \begin{figure}[t]
  \centering  
  \includegraphics[width=0.23\textwidth]{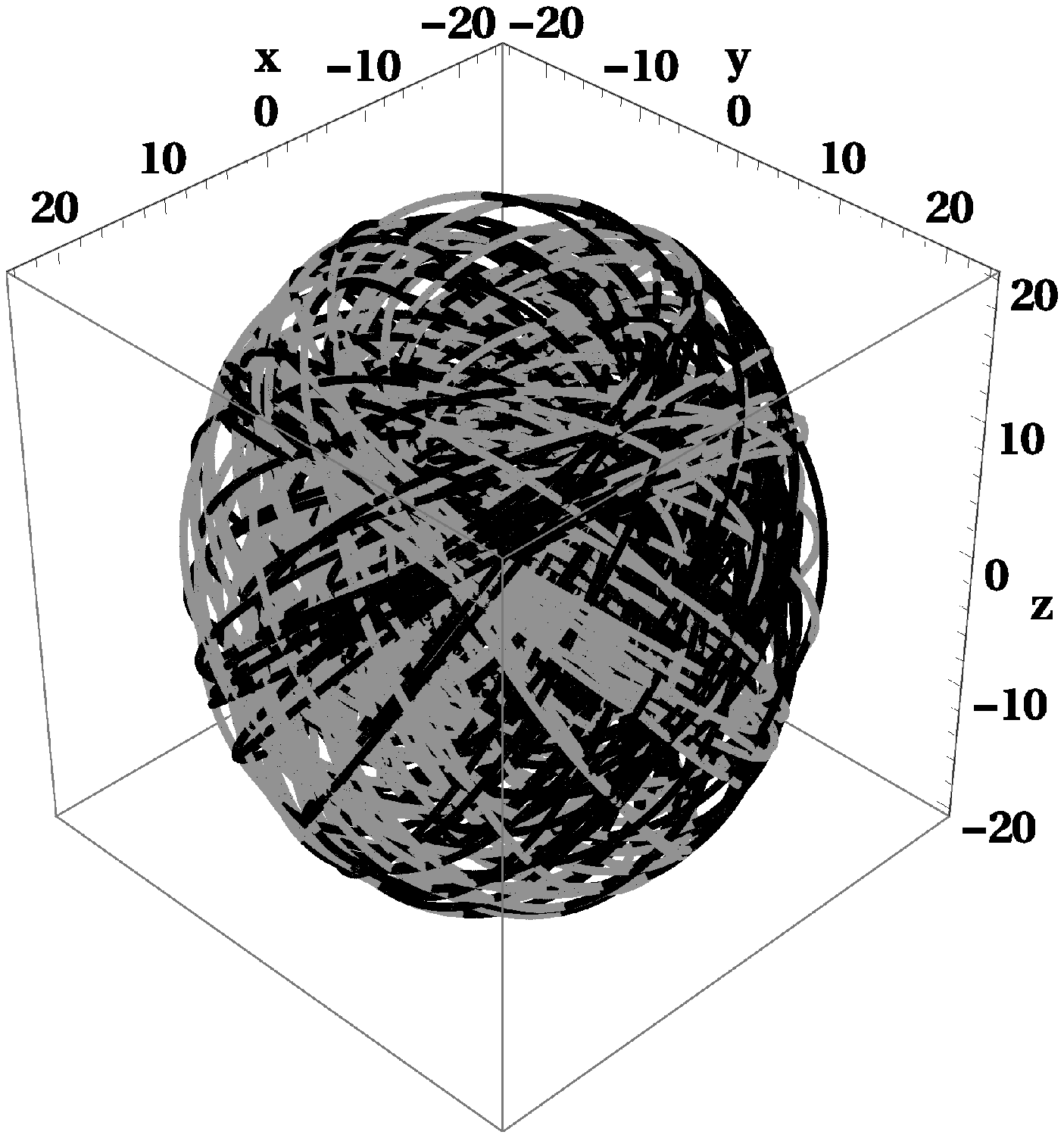} 
  \includegraphics[width=0.23\textwidth]{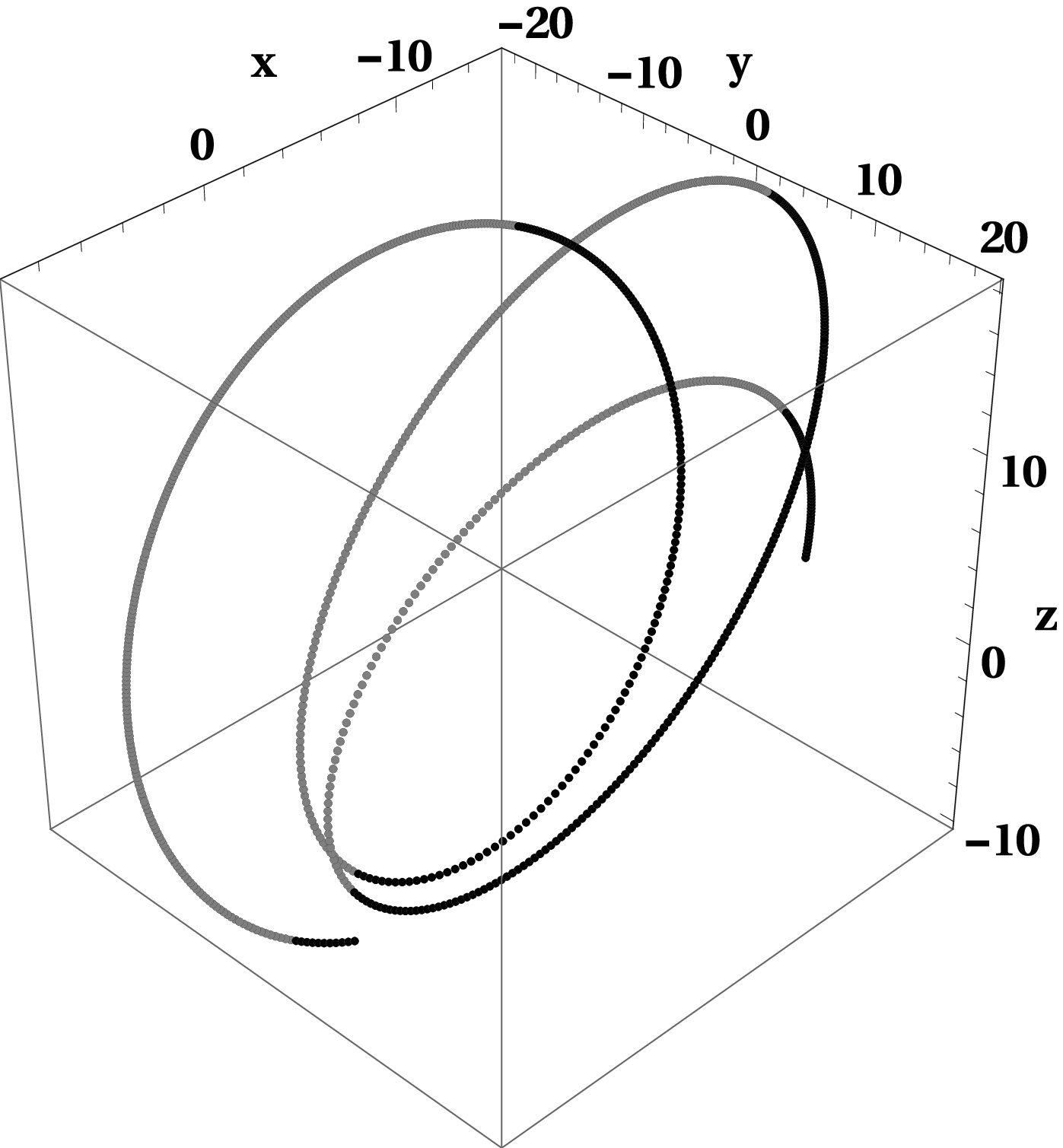} 
  \\
  \caption{Two orbits following the MPD equations with TD SSC are depicted in
  the configuration space. The orbits share the same initial conditions
  $r=10$, $\theta=\pi/2$, $u^r=0.1$, $S^r=0.1~S$, $S^\theta=0.01~S$ and constants
  of motion $m=1$, $S=0.9$, $E=0.97$ $J_z=3$.
  The black curve shows the orbit for which the affine parameter is the proper
  time, while the gray curve shows the orbit for which the affine parameter is 
  defined by the constraint~\eqref{eq:Ehlers}.
  The left panel shows the evolution of the orbits for $0\le \chi \le 10^5$,
  while the right for $0.99\times10^5 \le \chi \le 10^5$.
  } 
  \label{fig:3D}
  \end{figure}
   
   To show whether the under discussion time constraints reproduce the same
   worldline or not, initial conditions leading to a generic non-equatorial orbit
   has to be chosen. Such initial conditions produce the orbits shown in
   Fig.~\ref{fig:3D}. These orbits cover a non-zero width spheroidal shell around
   the central Kerr black hole (left panel). The pseudocartesian coordinates
   $(x,~y,~z)$ used in Fig.~\ref{fig:3D} relate to the BL ordinates as follows
    \begin{eqnarray}
     x &=& r \cos\phi \sin\theta\quad,\nonumber \\
     y &=& r \sin\phi \sin\theta\quad,\nonumber \\
     z &=& r \cos\theta\quad.\label{eq:CarCoor}
    \end{eqnarray} 
   The orbits evolve in a non-trivial manner; examples of trivial motion is
   a circular or a radial orbit. However, the orbits appear to follow exactly
   the same paths until the end (right panel of Fig.~\ref{fig:3D}). This implies
   that the two time constraints reproduce the same worldline. 
   
  \begin{figure}[t]
  \centering  
  {\includegraphics[width=0.35\textwidth]{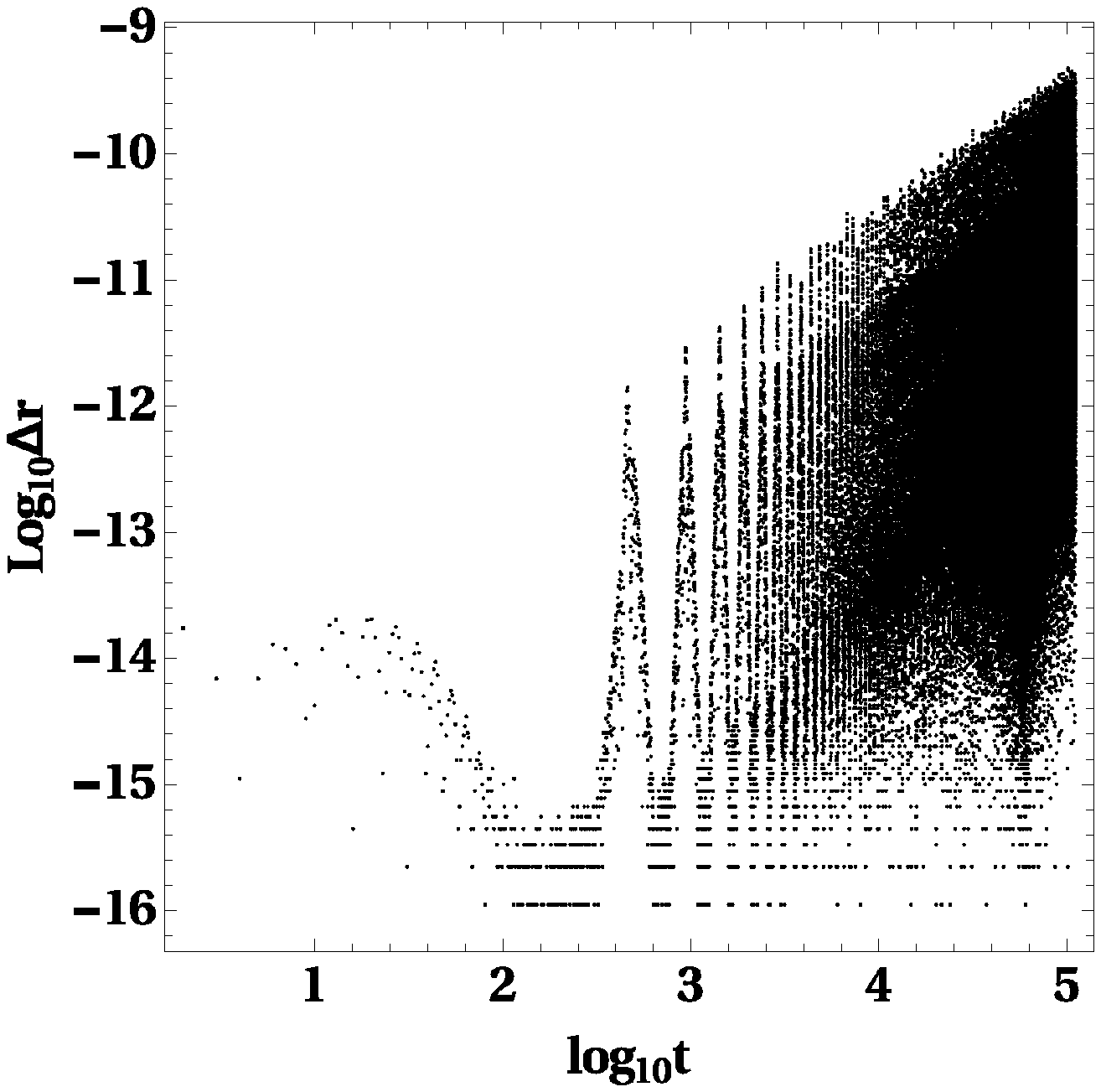} 
  \includegraphics[width=0.35\textwidth]{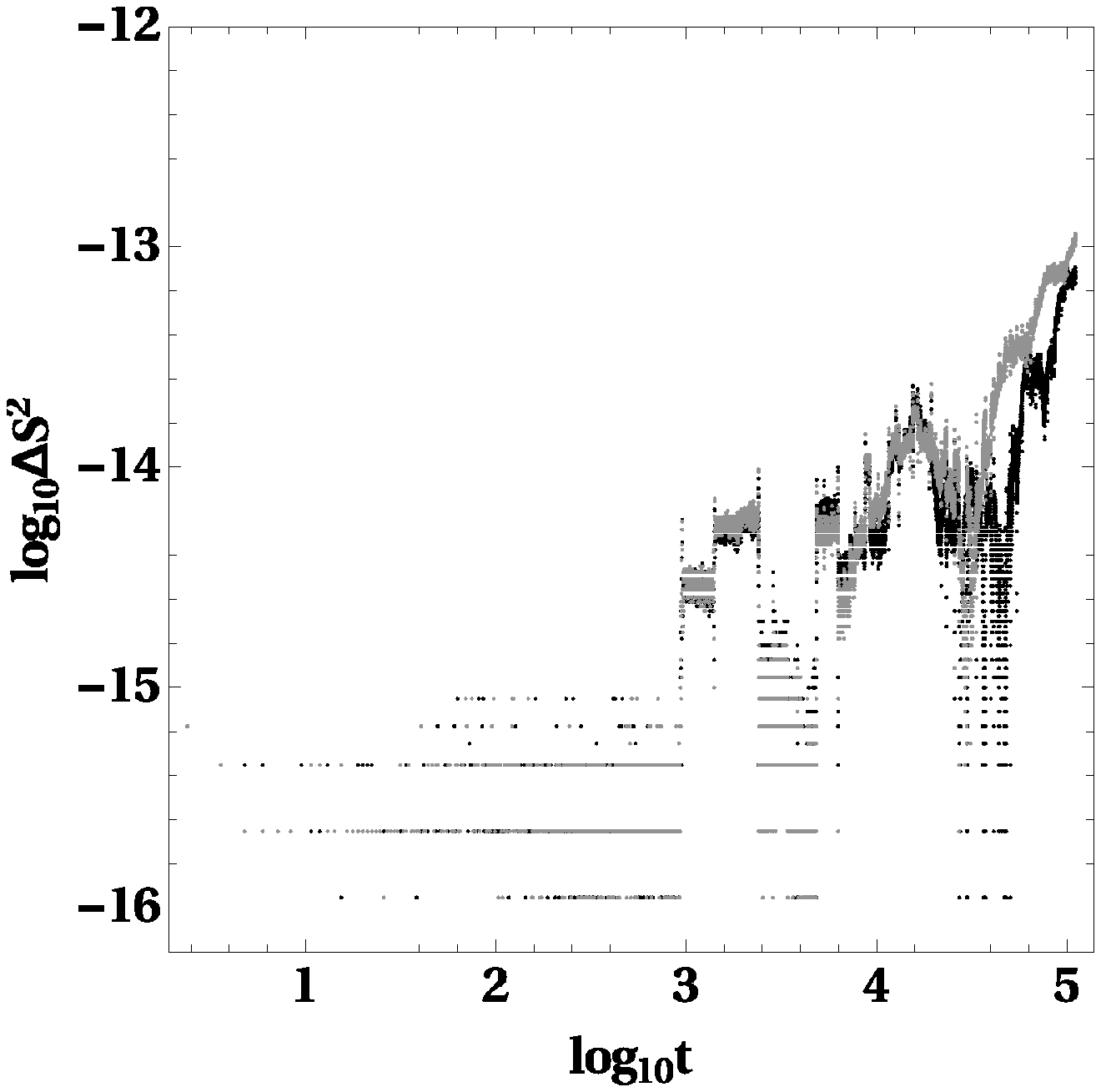} }
  \\
  \caption{Top panel: The relative radial difference $\Delta r$ of the two
  orbits shown in Fig.~\ref{fig:3D} as a function of the coordinate time $t$.
  Bottom panel: The relative error  in the spin measure~\eqref{eq:SpinMeas}
  $\Delta S^2$ of the two orbits shown in Fig.~\ref{fig:3D} as a function of $t$. 
  The colors denote the same as in Fig.~\ref{fig:3D}, i.e.
  black denotes the evolution using the proper time, while gray using the time
  constraint~\eqref{eq:Ehlers}.
  } 
  \label{fig:RelDif}
  \end{figure}  
  
  \begin{figure}[t]
  \centering  
   {\includegraphics[width=0.35\textwidth]{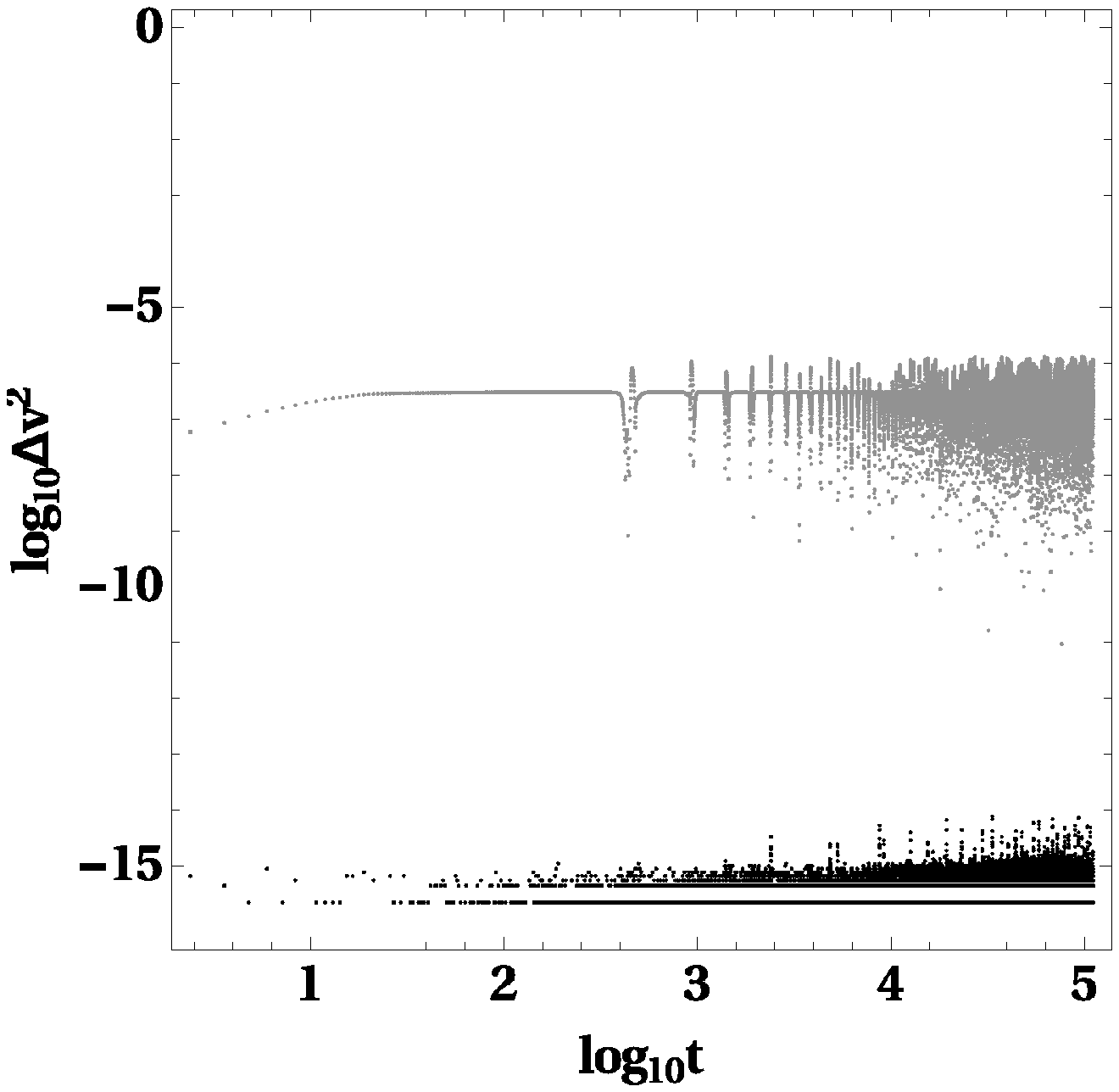}
     \includegraphics[width=0.35\textwidth]{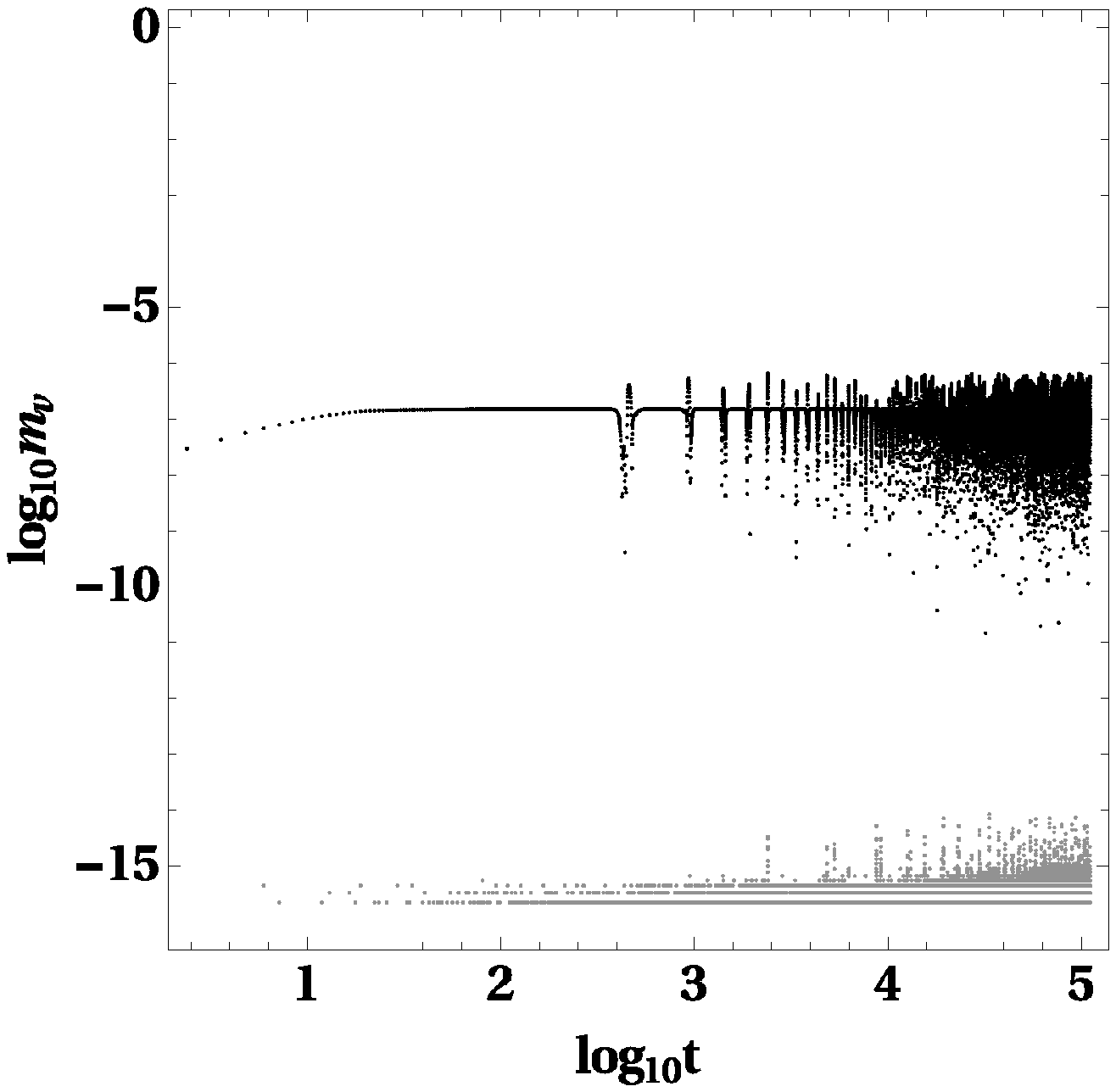} }
  \\
  \caption{Top panel: The relative error/difference (black/gray)
   $\displaystyle\Delta v^2=\left|1-\frac{v^2(t)}{v^2}\right|$
  in the value of Eq.~\eqref{eq:4VelCon} as the two orbits evolve with respect 
  to the coordinate time $t$, where $v^2(t)$ is the numerical value of 
  Eq.~\eqref{eq:4VelCon} calculated at time $t$ and $v^2$ is the initial value.
  Bottom panel: The relative difference/error (black/gray)
    $\displaystyle\Delta m_v=\left|1-\frac{m_v(t)}{m_v}\right|$
  in the value of Eq.~\eqref{eq:VelMass} as the two orbits evolve with respect 
  the coordinate time $t$, where $m_v(t)$ is the numerical value of 
  Eq.~\eqref{eq:VelMass} calculated at time $t$ and $m_v$ is the initial value.
  The colors denote the same as in Fig.~\ref{fig:3D}, i.e.
  black denotes the evolution using the proper time, while gray using the time
  constraint~\eqref{eq:Ehlers}.  
  } 
  \label{fig:ErDif}
  \end{figure}  
   
   To ensure that what is shown in Fig.~\ref{fig:3D} is not just an optical
   artifact, in the top panel of Fig.~\ref{fig:RelDif} is displayed the relative
   radial difference between the two orbits
   \begin{align}\label{eq:RelRad}
    \Delta r=\left|1-\frac{r_\sigma(t)}{r_\tau(t)}\right|
   \end{align}
   as a function of the coordinate time $t$. $r_\tau$ denotes the radial
   component of the orbit evolved using the proper time, while $r_\sigma$ denotes
   the radial component of the orbit using the affine parameter $\sigma$ defined
   by the constraint~\eqref{eq:Ehlers}. The coordinate time introduces a third
   observer at infinity with his own clock.
   This clock provides a common time by which the orbits can be compared.
   The top panel of Fig.~\ref{fig:RelDif} shows that the discrepancies in
   the radial component start being at the level of the computational accuracy, which
   is double precision, and after $t\sim 10^3$ they appear to drift away on average
   linearly. This drift resembles Fig.~11 in Ref.~\cite{LGSK}, where the integration
   scheme of s-stage Gauss Runge-Kutta used in this work was tested, and a similar
   drift was assigned to the interpolation used in the scheme. Actually, in order
   to produce the top panel of Fig.~\ref{fig:RelDif} interpolation was employed
   to get from a two component functional $\{r_\chi(\chi),t(\chi)\}$ to the function
   $r_\chi(t)$, since both orbits were computed using their respective affine
   parameters $(\chi=\sigma,~\tau$). Moreover, the bottom panel of
   Fig.~\ref{fig:RelDif} shows that the relative error of the spin measure
   \begin{align}\label{eq:RelErSpin}
     \Delta S^2=\left|1-\frac{S^2(t)}{S^2}\right|
   \end{align}
   increases on average linearly after $t\sim 10^3$ as well. $S^2(t)$ denotes
   the numerically computed value of $S^2$ at time $t$, while $S^2$ denotes the
   initial value of the spin. In few words, the drift between the two orbits
   shown in the top panel of Fig.~\ref{fig:RelDif} arises for numerical reasons, 
   and the two orbits reproduce the same worldline up to numerical accuracy.
  
  The fact that we do not see this drift in Fig.~\ref{fig:ErDif} for the
  four-velocity conservation~\eqref{eq:4VelCon} in the case of the proper time
  (black dots, top panel) and for the mass $m_v$ in the case of the affine
  parameter $\sigma$ (gray dots, bottom panel) is that at each step
  quantities $v^2$ and $m_v$ are normalized in order to compute the velocity
  through the relation~\eqref{eq:v_p_TUL}. Namely, for the proper time
  $v^2$ is kept equal to $1$, while for $\sigma$ $m_v$ is kept equal to $m$.
  Thus, it is no wonder why the relative errors of four-velocity conservation
  $v^2=1$ in the first case and of the mass $m_v=1$ in the second case stay at
  the computational accuracy level for so long. An interesting aspect of
  Fig.~\ref{fig:ErDif} is the evolution of the relative difference between
  the initial value of $v^2$ and the value of $v^2$ at time $t$ for the affine
  parameter $\sigma$ (gray dots, top panel), and of the relative
  difference between the initial value of $m_v$ and the value of $m_v$ at time
  $t$ for the proper time (black dots, bottom panel). The respective curves
  of the above two relative differences are practically identical, even the
  oscillations during the evolution take place at the same time. These curves
  provide a numerical example of the analysis provided in Sec.~\ref{sec:TDSSC}
  and show that the orbit does not belong to the special case for which
  $p^\mu||v^\mu$.
  
  It is notable that the phase space of the system does not change its
  dimensionality for the two time parameterization choices, i.e. the number of
  the constants of motion the same for $\tau$ and $\sigma$. Namely, in the case
  of the proper time the four-velocity~\eqref{eq:4VelCon} is preserved and the
  mass~\eqref{eq:VelMass} is not, and for the affine parameter the preservation
  is vice versa. If the number of constants was not the same, then this would
  imply that the two affine parameter choices alter the nature of the MPD
  equations and this choice is not just a gauge.
  
  \section{Conclusions} \label{sec:Concl}
   
  This article revisited relations derived from the Mathisson-Papapetrou-Dixon
  equations without specifying the affine parameter nor the spin supplementary
  condition. Next, the proper time choice versus the affine parameter choice
  introduced in \cite{Ehlers77} were discussed in the case of the
  Mathisson-Pirani SSC, the Tulczyjew-Dixon SSC and the Ohashi-Kyrian-Semer\'{a}k
  SSC, and the implications of this choice were analyzed. 
  
  In particular, it was found that under OKS SSC the affine parameters are
  identical, while for the MP SSC the choice of the affine parameter affects the
  preservation of the mass $m_v$. Namely, for the proper time choice $m_v$ is a 
  constant of motion, while for the Ref.~\cite{Ehlers77} choice the quantity
  ${m_v}^2/v^2$ is preserved instead.
  
  The TD SSC was not only approached analytically, but also numerically. The
  analytical approach focused on the implications brought by the fact that
  $m=m_v$. The numerical approach proved that the affine parameter choices
  $\tau$ and $\sigma$ are just a gauge choice, since both reproduce the same
  worldline when the MPD equations are evolved from the same initial conditions.

\begin{acknowledgments}
 G.L.-G. acknowledges the support from Grant No. GACR-17-06962Y and thanks
 O.~Semer\'{a}k for useful discussions.
\end{acknowledgments}

\end{document}